\input{aipcheck}

\documentclass[
    ,final            
  ]
  {aipproc}

\layoutstyle{6x9}

\begin{document}

\title{Understanding the Oldest White Dwarfs: Atmospheres of Cool WDs as Extreme Physics Laboratories
}

\classification{97.20.Rp,31.15A-,62.50.-p,97.10.Tk}
\keywords      {white dwarfs -- stars: atmospheres}


\author{Piotr M. Kowalski}{
  address={Helmholtz Centre Potsdam - GFZ German Research Centre for Geosciences, Telegrafenberg, 14473 Potsdam, Germany}
}

\begin{abstract}
Reliable modeling of the atmospheres of cool white dwarfs is crucial for understanding the atmospheric evolution of these 
stars and for accurate white dwarfs cosmochronology. 
Over the last decade {\it ab initio} modeling entered many research fields 
and has been successful in predicting properties of various materials under extreme conditions.
In many cases the investigated physical 
regimes are difficult or even impossible to access by experimental methods, and first principles quantum mechanical calculations 
are the only tools available for investigation. Using modern methods of computational chemistry and physics we investigate the atmospheres of helium-rich, old white dwarfs.
Such atmospheres reach extreme, fluid like densities (up to grams per cm$^3$) and represent an excellent laboratory for high temperature and pressure physics and chemistry. 
We show our results for the stability and opacity of $\rm H^-$ and $\rm C_2$ in dense helium and the implications of our work for understanding cool white dwarfs.

\end{abstract}

\maketitle


\section{Introduction}

\begin{figure}
\resizebox{\hsize}{!}{\rotatebox{270}{\includegraphics{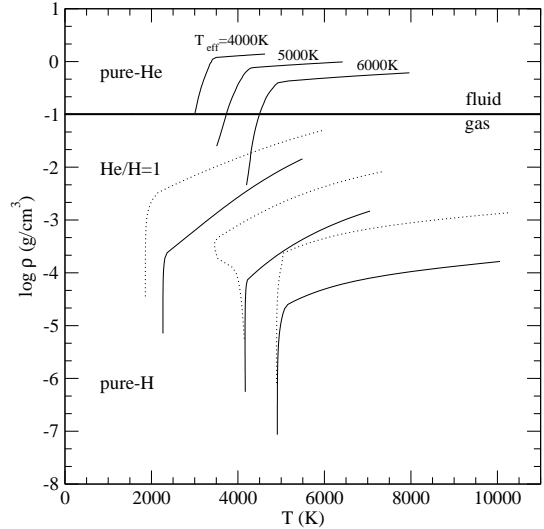}}}
  \caption{The $\rho-T$ atmosphere profiles of cool white dwarf atmosphere models of various H/He compositions and effective temperatures. 
  The horizontal line indicates the gas-fluid transition region at $\sim 0.1\rm\,g/cm^3$. \label{F1}}
\end{figure}

Cool white dwarf stars are among the oldest stars in our Galaxy. Having depleted their 
nuclear energy sources these stellar remnants simply cool-off releasing the internal heat to space. 
The coolest stars have been cooling for billions of years and the properties of their atmospheres
tells us about the stellar evolution history in our Galaxy.
Atmospheres of cool white dwarfs are not only of interest to astrophysicists but are excellent 
research objects for condensed matter scientists. Density-temperature profiles of cool white dwarfs atmospheres
are given in figure \ref{F1}. As helium is less transparent than hydrogen, helium-rich atmospheres
reach extreme, fluid-like conditions ($\rm \rho$ of a few $\rm g/cm^3$), under which the strong interparticle interactions affect 
atmospheric chemistry and physics.
They represent an excellent dense physics laboratory for investigation of properties of highly compressed, warm matter in 
the gas-plasma transition region.

\section{{\it Ab initio} investigations}

We aim to improve white dwarf atmosphere models by combining methods
of quantum mechanics and the theory of fluids \citep{KSM07,K10}. Because of the difficulties in describing 
the properties of fluid-like H/He mixture, until recently the atmosphere models were constructed using 
the ideal gas approximations. Improvements in computational hardware and methods
have allowed for computation and simulations of matter under extreme conditions.
This allows for derivation of more reliable input physics and chemistry for the models.
In our investigation we use density functional theory, which is most common quantum method used 
to treat many-particle systems. Most recently, we have investigated the properties 
of the negative hydrogen ion and molecular carbon.

\subsection{$H^-$ in dense helium}

\begin{figure}
\resizebox{\hsize}{!}{\rotatebox{270}{\includegraphics[width=2in]{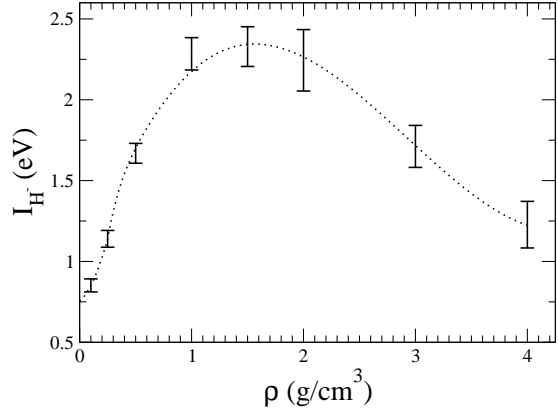}}}
  \caption{The ionization energy of the negative hydrogen ion as a function of helium density.
  \label{F2}}
\end{figure}

The stability of the negative hydrogen ion in dense helium is a test case problem in modeling of 
dense, fluid-like  H/He mixtures. \citet{Bergeron97} postulated that $\rm H^-$ undergoes pressure ionization in dense helium, 
which could explain the failure to detect the $\rm H^-$ bound-free opacity in 
the spectra of cool, helium-rich stars, whose atmospheres are expected to be enriched with hydrogen accreted from the ISM.
Our calculations show a different picture. The negative hydrogen ion becomes more stable in dense 
helium and does not undergo pressure ionization up to a density of $\rm 4 \, g/cm^3$.
Its ionization energy is given in figure \ref{F2}. It increases with density up to $\rm \rho\sim2 \, g/cm^3$,
and then gradually decreases at higher densities. The negative hydrogen ion probably pressure ionizes, but at densities much higher than
that found in the cool white dwarf atmospheres. The increase in the $\rm H^-$ ionization energy with helium density 
should produce a shift of the $\rm H^-$ bound-free absorption edge from the near-IR to the shorter wavelengths,
making this absorption mechanism invisible in the near-IR spectra of these stars.

\subsection{$C_2$ in dense helium}
\begin{figure}
\resizebox{\hsize}{!}{\rotatebox{270}{\includegraphics{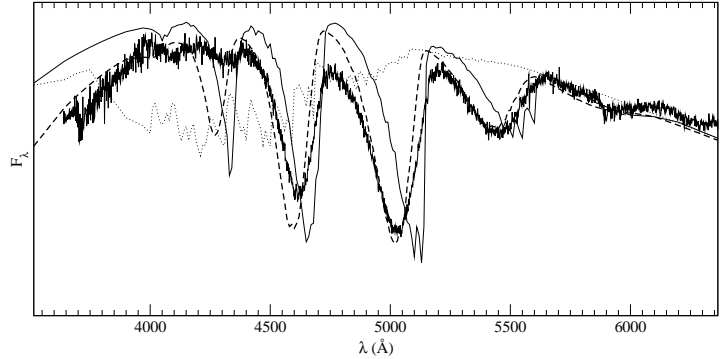}}}
  \caption{Fits to the optical spectrum of the DQp white dwarf LHS290 (thick solid line).
All the fits are with $T_{\rm eff}\rm=5800 \, K$ and log g=8 (cgs).
The parameters of the fits are: H/He=0, $\rm C/He=1.25\cdot 10^{-7}$, and $\Delta T_{\rm e}=1.6\, \rho_{\rm He} \rm \, eV$ (solid line); 
H/He=0, $\rm C/He=1.25\cdot 10^{-7}$, $\Delta T_{\rm e}=0\rm \, eV$ (dotted line); mixed composition $\rm He/H$ model, $\Delta T_{\rm e}=1.6\, \rho_{\rm He} \rm \, eV$, 
$\rm C/He=10^{-6}$ and $\rm H/He=6.75 \cdot 10^{-3}$ (dashed line).
All synthetic spectra well match the entire known spectral energy distribution of the star. The models are described in \citet{K10}. \label{F3}}
\end{figure}

Peculiar DQ stars (DQp) show distorted Swan bands of $\rm C_2$, which we have modeled
as pressure-shifted by interactions with the surrounding dense helium fluid (Kowalski, these proceedings).
We then modeled the optical spectrum of DQp star LHS290 \citep{K10} by shifting the Swan bands by the amount calculated as a function of density
in the atmosphere ($\Delta T\rm_e=1.6 \,\rho\,(eV)$).
The result is given in figure \ref{F3}. When we assume a pure helium composition the distortion of the Swan spectrum is larger than observed. 
This is because the density in the region of band formation in pure-He/C model is $\rm \sim 0.4\,g/cm^3$, which is 
much larger than the density required to produce the $\rm C_2$ bands distortions that would match the optical spectrum of this star ($\rho\sim \rm 0.05 \, g/cm^3$). 
The photospheric density is lower when the hydrogen is present in the atmosphere, resulting in smaller shifts. A modest amount of hydrogen (H/He=$6.75 \cdot10^{-3}$)
is required to reproduce the optical spectrum of LHS290. This finding is not unexpected as the derived H/He atmospheric composition is consistent with 
the H/He abundance derived by \citet{KB10} for the DQp star J1442+4013 of similar effective temperature ($T\rm_{eff}=5737\,K$, H/He=$2.1 \cdot10^{-3}$).

\section{Summary}
Because of the extreme, fluid-like conditions found inside their atmospheres the cool white dwarfs are ideal objects
for investigation of properties of warm, dense, H/He-rich matter. 
We apply modern methods of computational quantum mechanics to address interesting problems in condensed matter science and astrophysics.
We address the problem of stability of the negative hydrogen ion and the properties of molecular carbon in fluid-like dense helium. At odds with previous speculations, we found that $\rm H^-$ remains stable 
in dense helium under conditions of cool white dwarfs atmospheres, however its bound-free opacity edge shifts to the optical wavelengths,
removing this important opacity source from the near-IR. Investigation of $\rm C_2$ in dense helium shows that in cool white dwarfs
the optical spectral bands of molecular carbon should show a blueward shift as a result of
the increase in the electronic transition energy between states involved in the $\rm C_2$ Swan transition. This explains the origin of the DQp stars. 
The observed value of the shifts in the DQp stars indicates that the atmospheres of DQp stars are less dense than predicted by the pure-He/C models, 
implying that the atmospheres of these stars are probably polluted by hydrogen, which increases the opacity and decreases the density.
This finding requires confirmation by detailed analysis of the atmospheres of a larger sample of DQ/DQp stars.


\begin{theacknowledgments}
I wish to thank Didier Saumon for useful comments that improved the clarity of this paper and Sandro Jahn 
and Molecular modeling of geomaterials group at GFZ Potsdam for financial support.
\end{theacknowledgments}



\bibliographystyle{aipproc}   

\bibliography{sample}

\IfFileExists{\jobname.bbl}{}
 {\typeout{}
  \typeout{******************************************}
  \typeout{** Please run "bibtex \jobname" to optain}
  \typeout{** the bibliography and then re-run LaTeX}
  \typeout{** twice to fix the references!}
  \typeout{******************************************}
  \typeout{}
 }

\end{document}